# INVARIANT ALGEBRAIC SLICING OF THE SPACETIME[*]

C. BONA and J. STELA
*Departament de Fisica, Universitat de les Illes Balears*
*E-07071 Palma de Mallorca, SPAIN*

and

J. MASSO and E. SEIDEL
*National Center for Supercomputer Applications, 605 East Springfield Avenue*
*Champaign, IL 61280, USA*

## ABSTRACT

Using the momentum constraint, the standard evolution system is written in a fully first order form. The class of first order invariant algebraic slicing conditions is considered. The full set of characteristic fields is explicitly given. Characteristic speeds associated to the gauge dependent eigenfields (gauge speeds) are related to light speed.

## 1. Introduction

It is well known that Einstein's field equations can be decomposed into two sets: the evolution system and the constraints. There is no ambiguity in defining what we call constraint equations: this means equations in which there are no second time derivatives. Energy and momentum constraints are easily singled out in that way.

The term evolution system refers to the set of remaining equations. This is just a matter of choice, because an evolution equation plus a constraint leads to another evolution equation with the same physical solutions (the ones obtained from initial data which satisfy the constraints). The standard choice is to take the space components of the Ricci tensor, but one could choose instead the space components of the Einstein tensor or any other combination obtained by using the constraints in a suitable way, as we will actually do in the present work.

As it is well known, the freedom to choose the cordinate gauge allows one to complete the evolution system in many different ways and this can lead to many different systems of equations, each one with its own structure. Nevertheless, we know that the physical solutions belong to the solution space intersection of all these systems so that they are physically equivalent. However, this is not of much help in Numerical Relativity, where hyperbolic systems play an outstanding role. They allow one to apply the powerful methods of Computational Fluid Dynamics: there

---

is no need for building (and testing) new 'ad hoc' codes, one can just apply what it is known to work and concentrate himself in the physics of the problem considered.

In this work we will look for hyperbolic evolution systems for a wide class of coordinate gauges. We shall consider first order slicing conditions which are invariant under (time dependent) transformations of the space coordinates. We shall take for simplicity a vacuum spacetime and we will use a zero shift vector, but most of the results apply also to the general case.

## 2. Evolution Equations

The standard 3+1 evolution system is given in the vacuum case by:

$$\partial_t \gamma_{ij} = -2\alpha \ K_{ij} \tag{1}$$

$$\partial_t K_{ij} = -\alpha_{;ij} + \alpha \ [R^{(3)}_{ij} - 2K^2_{ij} + tr \ K \ K_{ij}] \tag{2}$$

This well-known system is of first order in time, but second order in space. To obtain a system which is also of first order in space, we will introduce auxiliary variables which correspond to the space derivatives,

$$A_k = \partial_k ln \ \alpha, \quad D_{kij} = 1/2 \ \partial_k \gamma_{ij}. \tag{3}$$

One could then simply insert these quantites into the standard ADM equations to obtain a first order system. However, doing so blindly does not have any particular advantage. As we show, a careful choice of variables will transform the equations into a special flux conservative, hyperbolic form that is especially suited to numerical treatment.

In particular, the evolution system (2) can then be written as

$$\partial_t K_{ij} + \partial_r (\lambda^r_{ij}) = \alpha \ S_{ij} \ . \tag{4}$$

where the terms $\lambda^k_{ij}$ are given by

$$\lambda^k_{ij} = D^k_{ij} + 1/2 \ \delta^k_i \ (A_j + 2 V_j - D_{jr}^{\ r}) + 1/2 \ \delta^k_j \ (A_i + 2 V_i - D_{ir}^{\ r}), \tag{5}$$

and we have noted for short

$$V_k = D_{kr}^{\ r} - D^r_{rk} \ . \tag{6}$$

$S_{ij}$ is a source term involving only the fields themselves and not their derivatives:

$$\begin{aligned} S_{ij} &= -2K_i^{\ k}K_{kj} + tr \ K \ K_{ij} - \Gamma_{ikr}\Gamma_j^{\ kr} + 4D_{kri}D^{kr}_{\ \ j} + \Gamma^k_{\ kr}\Gamma^r_{\ ij} \\ &- (2D^{kr}_{\ \ k} - A^r)(D_{ijr} + D_{jir}) + A_i(V_j - 1/2 \ D_{jk}^{\ \ k}) + A_j(V_i - 1/2 \ D_{ik}^{\ \ k}) \ . \end{aligned} \tag{7}$$

It is clear that one needs also to evolve the space derivatives. The simplest way of doing so is just to take the time derivative of (3) and interchange the order of space and time derivatives:

$$\begin{aligned} \partial_t A_k + \partial_k (\alpha \ Q) &= 0 \\ \partial_t D_{kij} + \partial_k (\alpha \ K_{ij}) &= 0 \end{aligned} \tag{8}$$

where we have noted
$$\partial_t ln\, \alpha = -\alpha\, Q, \qquad (9)$$
so that the choice of $Q$ will determine the slicing.

The vector $V_k$ is a very interesting quantity. One can compute its time derivative from (8) but we will use the momentum constraint to transform this equation into
$$\partial_t V_k = \alpha\, A_r\, (K^r_{\ k} - tr\, K\ \delta^r_k) + \alpha\, [(D_{kr}^{\ \ s} - 2 D_{rk}^{\ \ s}) - \delta^s_k\, (D_{rj}^{\ \ j} - 2 D^j_{\ jr})]\, K^r_{\ s}\ . \qquad (10)$$

We will consider $V_k$ as independent quantities to be evolved with equation (10). In that way, (6) becomes an algebraic constraint between $V$ and the spatial metric derivatives $D$ (the algebraic form of the momentum constraint). As we will see, this is crucial to ensure the hyperbolicity of the evolution system.

This set of equations (1,4,8-10) has the special form we seek. The entire non-linear system of the Einstein evolution equations is now written in the form of first order balance laws as
$$\partial_t u + \partial_i F^i(u) = S(u) \qquad (11)$$
which is familiar from many branches of physics. The structure of this system can be now easily investigated.

## 3. Invariant algebraic slicing

Note that we have *not* yet specified the slicing, because the proper time derivative $Q$ of the lapse function is yet to be given. We are interested in invariant slicing conditions. This means that the spacetime slicing provided by our coordinate condition must be invariant under any transformation of the space coordinates of every slice. We must use then slicing scalars, like $\alpha$, $Q$ or $tr\, K$ and their proper time derivatives (note that the shift "vector" does not behave as a slicing vector; it is a vector under time independent transformations only).

We also want to use an algebraic condition. If we restrict ourselves to zero order scalars, we can play only with $\alpha$ and we get either a geodesic slicing or one of its generalizations. If we allow also first order scalars, we get both $Q$ and $tr\, K$ into the play. The most general homogeneous algebraic condition is then
$$Q - f(\alpha)\, tr\, K = 0 \qquad (12)$$
where $f$ is an arbitrary function.

The geodesic slicing is then included as a subcase with $f = 0$. The maximal slicing is included also as a limiting case when $f$ diverges. The $f = 1$ case corresponds to the harmonic slicing. Another interesting case is the '1+log' slicing, obtained when $f = 1/\alpha$; it mimics maximal slicing near a singularity, when the lapse collapses to zero. The term '1+log' arises from the expression of $\alpha$ in terms of $\sqrt{\gamma}$ one obtains when integrating (12) in the eulerian case (zero shift vector). Note however that the invariance of (12) ensures that one can apply it to obtain the same slicing even with a nonzero shift vector.

## 4. The characteristic fields

Although the system of evolution equations (1,4, 8-10) is first order and flux conservative, we have not examined its hyperbolicity. When dealing with a first order system, one does not discuss hyperbolicity in general. One must first choose a fixed space direction (we will take the the $k$ coordinate axis) to discuss hyperbolicity by considering only space derivatives along the selected direction. This is a framework which does not match the usual one for second order equations, where there is no need for choosing a priori a direction and all derivatives are dealt with at a time.

It is clear that the normal lines will be characteristic (with zero speed). The corresponding characteristic fields are the lapse and the metric components, the combination appearing in (12), the vector $V$ and

$$A_k - f\, D_{kr}{}^r, \quad A_{k'}, \quad D_{k'ij} \quad (k' \neq k). \tag{13}$$

Also, the light cones are characteristic surfaces (speeds $\pm \alpha\sqrt{\gamma^{kk}}$). The corresponding eigenfields are

$$\sqrt{\gamma^{kk}}\, K_{ik'} \mp \lambda^k{}_{ik'}, \tag{14}$$

One gets also characteristic cones associated to the eigenfields

$$\sqrt{\gamma^{kk}}\, tr\, K \mp \sqrt{f}\, \lambda^r{}_{kr} \tag{15}$$

with characteristic speeds

$$\pm \alpha \sqrt{f\, \gamma^{kk}} \tag{16}$$

and we will call these 'gauge speeds' because their explicit dependence on the slicing condition. Gauge speed coincides with light speed only in the harmonic case. It becomes infinite for a maximal slicing, which can be considered as a limiting case of our condition (12).

It is clear that negative values of $f$ will lead to imaginary gauge speeds. Moreover, the set of eigenfields is complete only if $f \neq 0$. This means that the evolution system will be strictly hyperbolic iff $f > 0$. Note also that gauges with $f < 1$ will have poor singularity avoiding behaviour because gauge speed would be lower than light speed. Therefore, cases with $f \geq 1$ will look more appealing for most Numerical Relativity applications.

## 5. Acknowledgements

This work is supported by the Dirección General para la Investigación Científica y Técnica of Spain under project PB91-0335. J.M. acknowledges a Fellowship (P.F.P.I.) from Ministerio de Educación y Ciencia of Spain. We also acknowledge the support of NCSA and NSF grants PHY94-07882 and ASC93-18152.